\documentclass[10pt,twocolumn,superscriptaddress,aps,pra,balancelastpage]{revtex4-2}
\usepackage[utf8]{inputenc}
\usepackage{amsmath,amsfonts,amssymb,fullpage,color,graphicx,xcolor,physics}
\usepackage[colorlinks]{hyperref}

\usepackage{braket}
\newcommand{\gv}[1]{\ensuremath{\text{\boldmath$ #1 $}}}
\renewcommand{\abs}[1]{\left| #1 \right|} 

\newcommand{\mU}{{\mathcal{U}}}

\newcommand{\mM}{{\mathcal{M}}}

\newcommand{\mD}{{\mathcal{D}}}

\newcommand{\vs}{{\gv{s}}}
\newcommand{\vp}{{\gv{p}}}
\newcommand{\vq}{{\gv{q}}}

\newcommand{\bE}{{\mathbb{E}}}
\newcommand{\bV}{{\mathrm {Var}}}

\newcommand{\prob}{{\mathrm{Pr}}}

\renewcommand{\epsilon}{\varepsilon}
\newcommand{\appropto}{\mathrel{\vcenter{
  \offinterlineskip\halign{\hfil$##$\cr
    \propto\cr\noalign{\kern2pt}\sim\cr\noalign{\kern-2pt}}}}}

\let\baraccent=\= 
\renewcommand{\=}[1]{\stackrel{#1}{=}} 

\begin{document}
\title{Shadow Distillation: Quantum Error Mitigation with Classical Shadows for Near-Term Quantum Processors}
\author{Alireza Seif}
\thanks{These authors contributed equally to this work.}
\affiliation{Pritzker School of Molecular Engineering, University of Chicago, Chicago, IL 60637}
\author{Ze-Pei Cian}
\thanks{These authors contributed equally to this work.}
\affiliation{Department  of  Physics,  University  of  Maryland,  College  Park,  Maryland  20742,  USA}
\affiliation{Joint Quantum Institute, University of Maryland, College Park, Maryland 20742,
USA}
\affiliation{Center for Quantum Information and Computer Science, University of Maryland,
College Park, MD 20742, USA}
\author{Sisi Zhou}
\affiliation{Pritzker School of Molecular Engineering, University of Chicago, Chicago, IL 60637}
\affiliation{Institute for Quantum Information and Matter, California Institute of Technology, Pasadena, CA 91125}
\author{Senrui Chen}
\affiliation{Pritzker School of Molecular Engineering, University of Chicago, Chicago, IL 60637}
\author{Liang Jiang}
\affiliation{Pritzker School of Molecular Engineering, University of Chicago, Chicago, IL 60637}
\begin{abstract}
Mitigating errors in quantum information processing devices is especially important in the absence of fault tolerance. An effective method in suppressing state-preparation errors is using multiple copies to distill the ideal component from a noisy quantum state. Here, we use classical shadows and randomized measurements to circumvent the need for coherent access to multiple copies at an exponential cost. We study the scaling of resources using numerical simulations and find that the overhead is still favorable compared to full state tomography. We optimize measurement resources under realistic experimental constraints and apply our method to an experiment preparing Greenberger–Horne–Zeilinger (GHZ) state with trapped ions. In addition to improving stabilizer measurements, the analysis of the improved results reveals the nature of errors affecting the experiment. Hence, our results provide a directly applicable method for mitigating errors in near-term quantum computers. 
\end{abstract}
\maketitle

\section{Introduction}
One of the main obstacles in operating quantum information processing devices is extreme sensitivity to errors. In principle, these errors can be corrected using error-correcting codes~\cite{lidar2013quantum}. However, utilizing these codes in a fault-tolerant manner requires a hardware overhead that is pushing the limits of what experiments can achieve today~\cite{Egan2021}. Therefore, it is interesting to find ways to mitigate the effect of errors and extend the utility of current devices in the absence of fault tolerance. Recently, there have been several proposals for mitigating the effect of errors on estimating expectation values of observables in a quantum circuit~\cite{temme2017error,endo2018practical,czarnik2020error,strikis2020learning,lowe2021unified,kandala2019error}. These schemes work by acquiring the expectation value of an observable for different noise strengths (e.g., by changing gate time) and extrapolating them to find the expectation value at the zero-noise limit, or as shown in Refs.~\cite{strikis2020learning,czarnik2020error} by learning a correction scheme using circuits that are easy (e.g., Clifford circuits) to simulate and applying the learned correction procedure to general circuits. Additionally, there has been a new endeavor along the ideas of Ref.~\cite{peres1999error} to extract the state of interest from a noisy mixed state by using multiple copies of the noisy state~\cite{cai2021resource,cotler2019cooling,huggins2020virtual,huo2021dual,koczor2021exponential,xiong2021quantum,lowe2021unified}. 

At the same time, quantum devices are growing in size, and that increases the complexity of extracting information from the system. In particular, methods such as quantum state tomography have a complexity that grows exponentially with the system size. Recently, there have been proposals for efficient extraction of certain properties of a quantum system based on randomized measurements and classical shadows~\cite{huang2020predicting,paini2021estimating,chen2021robust}. Roughly speaking, these methods provide a way for estimating many linear functions of a quantum state with (quantum and classical) resources that scale efficiently with the system size. For nonlinear functions of the state, such as R\'enyi entropies or topological invariants, protocols based on randomized measurements have an exponential complexity, but are still advantageous compared to full state tomography~\cite{elben2019statistical,elben2020cross,rath2021importance, elben2020many,brydges2019probing,cian2021many,huang2020predicting}, making them a useful tool for probing near-term intermediate scale devices~\cite{preskill2018quantum}.

\begin{figure}
    \centering
    \includegraphics[width=\columnwidth]{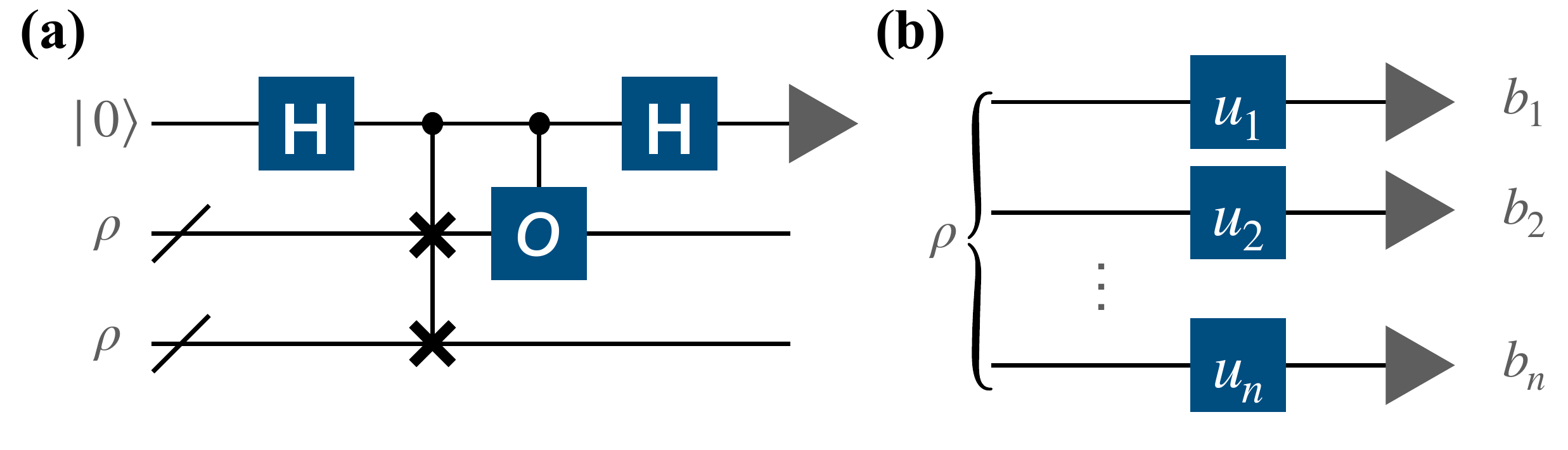}
    \caption{Schematic representation of error mitigation with multiple copies. (a) Performing an interferometry experiment with a controlled-SWAP on two copies of the state $\rho$ together with a controlled-$O$ operation on one of the copies enables measuring $\tr(O \rho^2)$. (b) The procedure in panel (a) can be replaced by randomized single-qubit measurements implemented by $u_i$ and post processing the results $b_i$. }
    \label{fig:schematic}
\end{figure}

In this work, we take advantage of the framework of randomized measurements and classical shadows and apply it to the problem of error mitigation. Specifically, we study error mitigation using multiple copies~\cite{huggins2020virtual,huo2021dual,koczor2021exponential,xiong2021quantum,lowe2021unified} and study the trade-off between quantum resources (such as two-qubit gates and coherent access to multiple copies of a state) and single-qubit randomized measurements~(see Fig.~\ref{fig:schematic}). We first explain the error mitigation framework and show how our protocol incorporates randomized measurements in this framework. We then provide a numerical analysis of the errors and resources and explore the trade-off between the number of measurement settings and the repetitions of each measurement. Finally, using the existing trapped-ion experimental data from Ref.~\cite{zhu2021cross}, we illustrate the application of our method in optimizing experimental resources for improving the measurements of  stabilizers of a 5-qubit Greenberger–Horne–Zeilinger (GHZ) state~\cite{greenberger1989going}. The success and shortcomings of our protocol, in this case, reveal valuable information about the nature of errors in the experiment. 

\section{Error mitigation using multiple copies}
We first review the scheme using multiple copies for suppressing errors in preparing a quantum state. Let $\ket{\psi}$ denote the ideal state that we are interested in preparing in an experiment. Due to experimental imperfections, we instead end up with $\rho = (1-\epsilon) \ketbra{\psi} + \epsilon \rho_{\rm{error}}$, where $0 < \epsilon \leq 1$ quantifies the strength of errors. We assume that $\rho_{\rm{error}}$ is a density matrix in a subspace orthogonal to $\ket{\psi}$, i.e., $\expval{\rho_{\rm{error}}}{\psi}=0$. Realistic noise in an experiment might differ from this model. In Supplementary Material we discuss the effectiveness of this scheme for various noise models. Now let us consider the task of estimating the expectation value of an observable $O$. Ideally, we  would like to extract $\expval{O}{\psi}$. However, because of the errors we obtain $\tr(O\rho)$. To reduce the errors in our estimate, one can instead calculate $\expval{O}_{(m)}=\tr(O \rho^m)/\tr(\rho^m)$, where $m$ is an integer, which is referred to as Virtual Distillation in the literature (see e.g., Refs.~\cite{koczor2021exponential,huggins2020virtual}). This scheme is effective if $\ket{\psi}$ is the dominant eigenvector of $\rho$, i.e., $1-\epsilon>\epsilon p_{\rm{max}}$ with $p_{\rm{\max}}$ being the largest eigenvector of $\rho_{\rm{error}}$, and  suppresses the errors exponentially in $m$~\cite{huggins2020virtual,koczor2021exponential} since
\begin{equation}
\begin{split}
\frac{\tr( O \rho^m)}{\tr(\rho^m)} &= \frac{\tr\{O [(1-\epsilon)^m\ketbra{\psi} + \epsilon^m \rho^m_{\rm{error}}]\}}{\tr[(1-\epsilon)^m\ketbra{\psi} + \epsilon^m \rho^m_{\rm{error}}]}\\&= \frac{(1-\epsilon)^m \expval{O}{\psi}+\epsilon^m \tr(O\rho_{\rm{error}}^m)}{(1-\epsilon)^m+\epsilon^m \tr(\rho_{\rm{error}}^m)}\\&\simeq \expval{O}{\psi}+f(O,\rho_{\rm{error}})\epsilon^m+\mathcal{O}(\epsilon^{m+1})
\end{split},
\end{equation}
where $f(O,\rho_{\rm{error}})=\tr(O\rho_{\rm{error}}^m)-\expval{O}{\psi}\tr(\rho_{\rm{error}}^m)$.
Hence, the access to $\rho^m$ enables suppressing errors exponentially in $m$. Previous works~\cite{cai2021resource,cotler2019cooling,huggins2020virtual,czarnik2021qubit,huo2021dual,koczor2021exponential,xiong2021quantum,lowe2021unified} have mostly considered using multiple copies and controlled permutations to prepare $\rho^m$, given access to $m$ copies of $\rho$. This is enabled by using the fact that $\tr(V^{(m)} \rho^{\otimes m})=\tr(\rho^m)$, where $V^{(m)}$ is a permutation operator acting as $V^{(m)}\ket{\psi_1}\ket{\psi_2}\dots\ket{\psi_m}=\ket{\psi_m}\ket{\psi_1}\dots\ket{\psi_{m-1}}$. Such schemes require the use of two-qubit gates between copies of the state $\rho$ stored in quantum registers. Note that while $m$ copies of $\rho$ are required for such a procedure, we only need coherent access to two copies at the same time~\cite{czarnik2021qubit}. Recently, there have been proposals to trade access to copies of the state (circuit width) with circuit depth using a dual-state scheme~\cite{cai2021resource,huo2021dual}. These methods eliminate the need for quantum operations between different copies of the state, which can be challenging in near-term devices~\cite{linke2018measuring}. However,
they require the knowledge of the unitary operator that prepares the state of interest and assume that the noise affecting the state and its dual are similar. The increased depth of the circuit can be problematic for the latter assumption in the presence of non-Markovian errors~\cite{hakoshima2021nonmarkovian}. 

\section{Shadow distillation}
In this work, we propose using the framework of randomized measurements and classical shadows to calculate $\tr(O \rho^m)$ and $\tr(\rho^m)$. Our method, which we refer to as Shadow Distillation (SD),  is useful for near-term devices, where control and circuit depth and width are limited and errors are large. Such an approach trades circuit size with sample complexity.

Specifically, let $\rho$ denote the state of interest on $n_q$ qubits. To measure the quantum state in $N_U$ random bases, we sample $N_U$ distinct combinations of random single-qubit rotations $ U = u_1 \otimes u_2 \otimes \dots \otimes u_N$ and append them to the circuit that is used to prepare $\rho$. Finally, we perform projective measurements on the computational basis. For each rotation setting $U$, the measurements are repeated $N_S$ shots. 

To infer the physical quantities from the randomized measurements, one can convert each measurement outcome to a classical snapshot of the state. For a  measurement with a random unitary $U = u_1 \otimes u_2 \otimes \dots \otimes u_N$ satisfying the 3-design property and a measurement outcome $|b\rangle = |b_1, b_2,.\dots , b_N\rangle$, the classical snapshot is of the form
\begin{equation}\label{eq:snapshot}
    {\hat{\rho}}_{U,b} = \otimes_{k=1}^{n_q} (3u_k^\dagger | b_k\rangle \langle b_k |u_k - I),
\end{equation}
where $I$ is the identity matrix on a single qubit. The collection of these snapshots is referred to as a classical shadow of the state~\cite{huang2020predicting}. The density matrix $\rho$ can be inferred from the classical shadow by averaging over $U$ and $b$, i.e., $\rho = \mathbb{E}_{U,b}(\hat{\rho}_{U,b})$. Therefore, one can directly infer the expectation value of an observable $O$ from its expectation value over each snapshot using
$\tr(O \rho) = \mathbb{E}_{U,b} [ \tr(O \hat{\rho}_{U,b})]$~\cite{huang2020predicting}. 
Physical quantities that are non-linear in the density matrix $\rho$, e.g., $\tr(O \rho^2)$, can be calculated through
$ \tr(O\rho^2) = \mathbb{E}_{U,b, U', b'}
    [\tr( V^{(2)} (O\hat{\rho}_{U,b}) \otimes \hat{\rho}_{U', b'})],$ 
where $V^{(2)}$ is the swap operator~\cite{huang2020predicting}. For certain choices of measurement bases, such as those corresponding to random Clifford operations and random Pauli measurements, the shadows can be stored and manipulated efficiently in a time and memory polynomial in $n_q$, $N_U$, and $N_S$~\cite{huang2020predicting}. 

Here, we focus on  second-order error mitigation  ($m=2$) with randomized single-qubit Pauli measurements. Specifically, let $\{U_j\}_{j=1}^{N_U}$ denote the $N_U$ sampled unitary operators from random local Clifford gates, and $\{\ket{b^{(i_j)}}_{i_j=1}^{N_S}\}$ denote the measurement outcomes of $N_S$ measurements fixing $U=U_j$. We then define $\hat\rho_j =\frac{1}{N_S}\sum_{i_j=1}^{N_S} \hat{\rho}_{U_j,b^{(i_j)}}$, which corresponds to the average snapshot~\eqref{eq:snapshot} for a fixed $U$. We denote our estimate of $\tr(O \rho^2)$ by  $\hat{o}_2$ given by~\cite{elben2020mixed}
\begin{equation}
\label{eq:shadow_average}
    \hat{o}_2 = \frac{1}{N_U(N_U-1)}\sum_{j\neq j'}\trace( {V^{(2)}} \hat\rho_j \otimes (O\hat\rho_{j'})),
\end{equation}
which is an unbiased estimator (see Supplementary Material). Note that setting $O=I$ results in an estimate of $\tr(\rho^2)$, which we denote by $\hat{s}_2$.  

In this way, $\hat{s}_2$ using $N_U N_S$ snapshots can be calculated in time $\mathcal{O}(\text{poly}(n) N_U^2 N_S^2)$. Moreover, $\hat{o}_2$ for operators $O$ that are products of single-qubit Pauli operators can be obtained with the same complexity~\cite{gottesman1998heisenberg}. Therefore, using classical shadows enables us to perform error mitigation for such operators using classical computational resources that scale polynomially with the number of samples and the number of qubits $n_q$. However, it should be noted that the number of samples required to achieve a given accuracy can depend on $n_q$. In fact, the sample complexity of estimating quantities nonlinear in the state $\rho$ can grow exponentially with system size~\cite{huang2020predicting,chen2021hierarchy}. In the following, we numerically investigate this scaling and show that for the case of $m=2$, $\expval{O}_{(2)}$ for Pauli observables, performs favorably compared to schemes based on full quantum state tomography. 

\section{Numerical investigation of error scaling}
We analyze the scaling of statistical errors in the estimation of $\expval{O}_{(2)}$ for Pauli observables with  measurement resources, $N_U$ and $N_S$, and the number of qubits $n_q$ using numerical simulations. To study the generic performance of the protocol, we first prepare random pure states under depolarization noise with strength $\epsilon$ 
\begin{eqnarray}
\rho_R = (1-\epsilon) |\psi_R\rangle \langle \psi_R| + \frac{\epsilon}{2^{n_q}-1} [I- |\psi_R\rangle \langle \psi_R|],
\label{eq:random mixed state}
\end{eqnarray}
where $0 < \epsilon \leq 1$, $|\psi_R\rangle = U_R|0\rangle$ and $U_R$ is a Haar random unitary operator. We then estimate $\tr(\rho_R^2)$ and $\tr(O \rho_R^2)$, denoted by $\hat{s}_2^{(R)}$ and $\hat{o}_2^{(R)}$, respectively,  using Eq. \eqref{eq:shadow_average}, by sampling $N_U$ random bases and $N_S$ shots. Let 
\begin{equation}
    \Delta^2_R = \left(\frac{\tr(O \rho_R^2)}{\tr(\rho_R^2)} - \frac{\hat{o}_2^{(R)}}{\hat{s}_2^{(R)}}\right)^2,
    \label{eq:shadow_estimation_error}
\end{equation}
denote the squared error of estimating $\expval{O}_{(2)}$ for the particular state $\rho_R$. In our simulations, we examine the mean squared error (MSE) $\Delta^2 = \frac{1}{N_R}\sum_R \overline{\Delta_R^2}$, over $N_R=100$ random choices of $U_R$. The overbar denotes the average taken over different realizations of measurements for each $U_R$ obtained by bootstrap sampling over $250$ instances, see Supplementary Material for more information on the bootstrap resampling techniques. We emphasize that $\Delta$ only captures errors of our SD scheme for estimating $\expval{O}_{(2)}$ and does not include the errors that are not corrected using this error mitigation procedure. The effectiveness of the error mitigation scheme has been studied in other works, see e.g., \cite{cai2021resource,huggins2020virtual,huo2021dual,koczor2021exponential,xiong2021quantum}. We  discuss that aspect in the discussion of our results for the trapped-ion experiment.   

\begin{figure}[t]
    \includegraphics[width=\columnwidth]{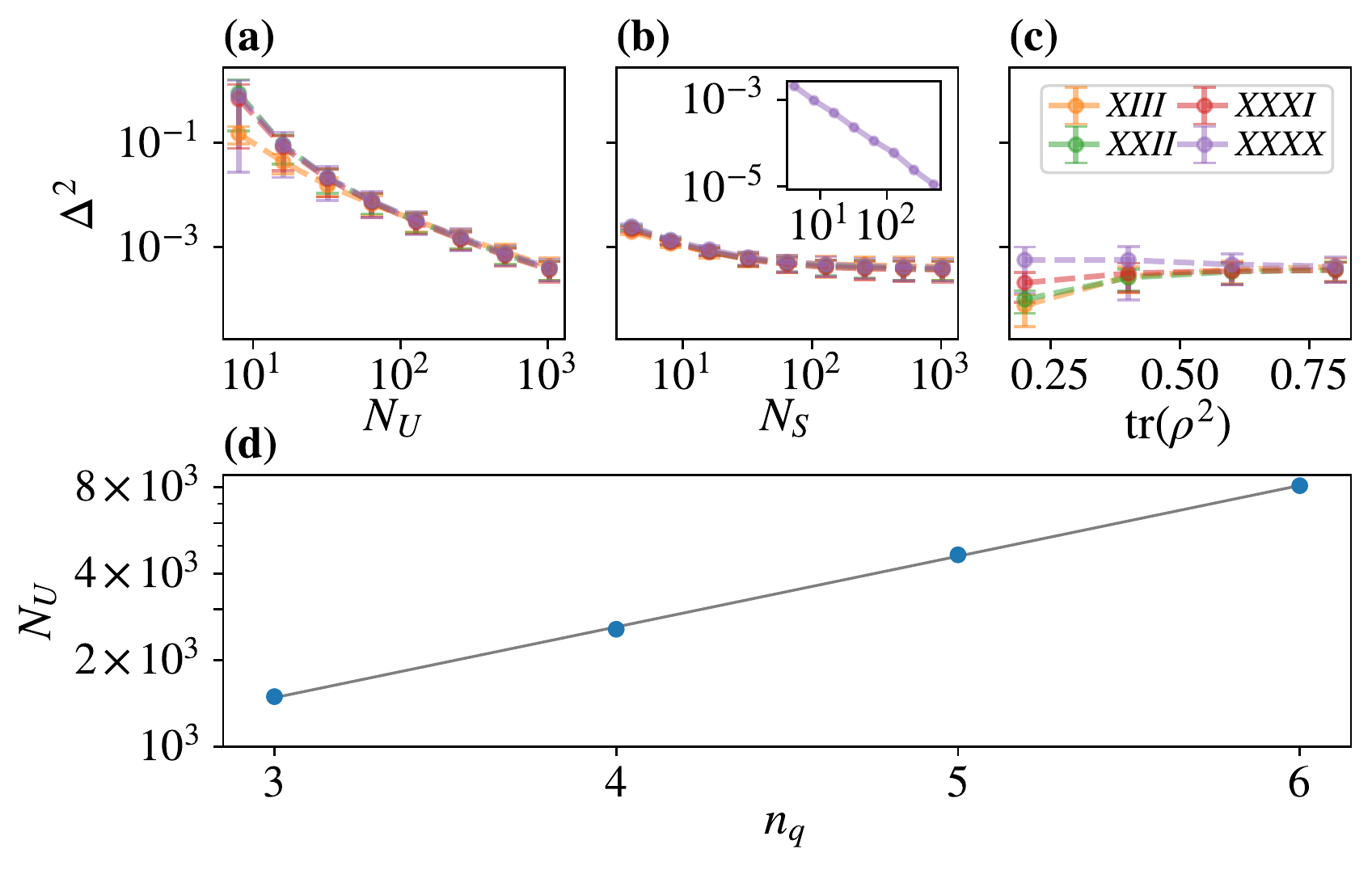}
    \caption{Scaling of the mean squared error $\Delta^2$ for $n_q=4$ qubits and error strength $\epsilon=0.1$ with (a) the number of unitaries $N_U$ with a fixed number of shots $N_S = 1024$ (b) the number of shots with a fixed $N_U = 1024$. The inset shows the convergence to the final value with $N_S$. (c) The scaling of $\Delta^2$ with purity for $N_S=N_U=1024$. The legend indicates the choices for $O$ (see Eq.~\eqref{eq:shadow_estimation_error}) in panels (a)-(c). Error bars are standard deviation of $\Delta^2$ over 100 random states.  (d) The number of basis $N_U$ in order to reach mean squared error $\Delta^2$ versus number of qubits $n_q$ for $N_S = 1$. The solid lines are fitting curve $N_U = c2^{\gamma n_q}$. For $\Delta = 0.01$, $\gamma = 0.82$.   }
    \label{fig:all_scalings}
\end{figure}
Figure \ref{fig:all_scalings}(a) and (b) show the scaling of statistical error as function of $N_U$ and $N_S$ for various observables $O$ for $n_q = 4$ and $\epsilon=0.1$. We observe that $\Delta^2$ scales as $1/{N_U}$. Moreover, for a fixed value of $N_U$, it converges as $1/N_S$ to a constant determined by $N_U$. We also observe a fast convergence of $\Delta^2$ to a constant value determined by $N_U$ and $N_S$ as a function purity, $\tr(\rho^2)$, as shown in Fig.~\ref{fig:all_scalings}(c). Note that our estimator $\hat{o}_2/\hat{s}_2$ is, in general, a biased estimator for $\expval{O}_{(2)}$. Moreover, there is no closed form formula for the variance of $\hat{o}_2/\hat{s}_2$. In Supplementary Material we derive an analytical bound for ${\rm{Var}}(\hat{o}_2)$ and ${\rm{Var}}(\hat{s}_2)$. While these bounds do not directly translate to a bound on ${\rm{Var}}(\hat{o}_2/\hat{s}_2)$ they can still provide an intuition on the behavior of the errors and help us find empirical expressions for the scaling of errors. In fact, the scaling that we observe in Fig.~\ref{fig:all_scalings}(a)-(b) agrees with our bound for the variance of the numerator.  Additionally, errors in $\hat{s}_2$ can lead to large errors in estimating the ratio $\hat{o}_2/\hat{s}_2$ especially in the small $N_U$ regime. In these cases it might be beneficial to incorporate prior knowledge about the value of the purity $\tr(\rho^2)$ to reduce the errors. We further explore this idea in the Supplementary Material. We show that given a measurement of purity $s_2$, a prior guess for the value of the purity $\mu_0$, and a hyperparameter $\alpha$ that quantifies the confidence in our guess,  a modified estimator of the form $(s_2+\lambda \mu_0)/(1+\lambda)$, where $\lambda=\alpha/N_U$ can be obtained using Bayes' rule.  

Finally, in Fig.~\ref{fig:all_scalings}(d) we investigate the number of basis measurement $N_U$ required to reach a certain value of $\Delta^2$ as a function of number of qubits $n_q$ with $\epsilon=0.1$. We find that although $N_U$ scales exponentially with $n_q$, i.e., $N_U\sim 2^{\gamma n_q}$, the exponent $\gamma \approx 0.82$, which is favorable compared to  full quantum state tomography with $N_U\sim 3^{n_q}$~\cite{odonnell2016efficient}. Therefore, the scheme is favorable for the near-term regime, where we are pushing the boundaries of the classical simulability of quantum systems.

\section{Trapped ions experiment}
We illustrate the utility of our proposed SD method, by applying it to the existing data from an experiment with trapped-ion qubits~\cite{zhu2021cross}, see also Supplementary Material for more information on the experimental device. 

In the experiment, a 5-qubit GHZ state, i.e., $\ket{\psi_{\rm{GHZ}}}=\frac{1}{\sqrt{2}}(\ket{0}^{\otimes 5}+\ket{1}^{\otimes 5})$ is prepared. This is a stabilizer state with generators $\mathcal{G}=\{Z_1 Z_2, Z_2 Z_3, Z_3 Z_4, Z_4 Z_5, \prod_i X_i \}$, where we use $\prod_i X_i$ to denote $X_1 X_2 X_3 X_4 X_5$~\cite{gottesman1998heisenberg}. Ideally, for this state $\expval{O}=1$ for all $O \in \mathcal{G}$.
Because of experimental errors, the actual state $\tilde{\rho}_{{\rm{GHZ}}}$ differs from the ideal state and $\expval{O}\leq 1$. Here, we investigate how our proposed error mitigation technique can improve estimates of these expectation values. Note that these expectation values can then be used to estimate the fidelity of the GHZ state~\cite{flammia2011direct,desilva2011practical,kalev2019validating}. A practical consideration in this experiment is that performing measurements in different bases takes roughly 1000 times longer than repeating measurements in a fixed basis. Therefore, it is  interesting to explore the possibility of a trade-off between the $N_U$ and $N_S$ for a fixed measurement time. 

\begin{figure}
    \includegraphics[width=\columnwidth]{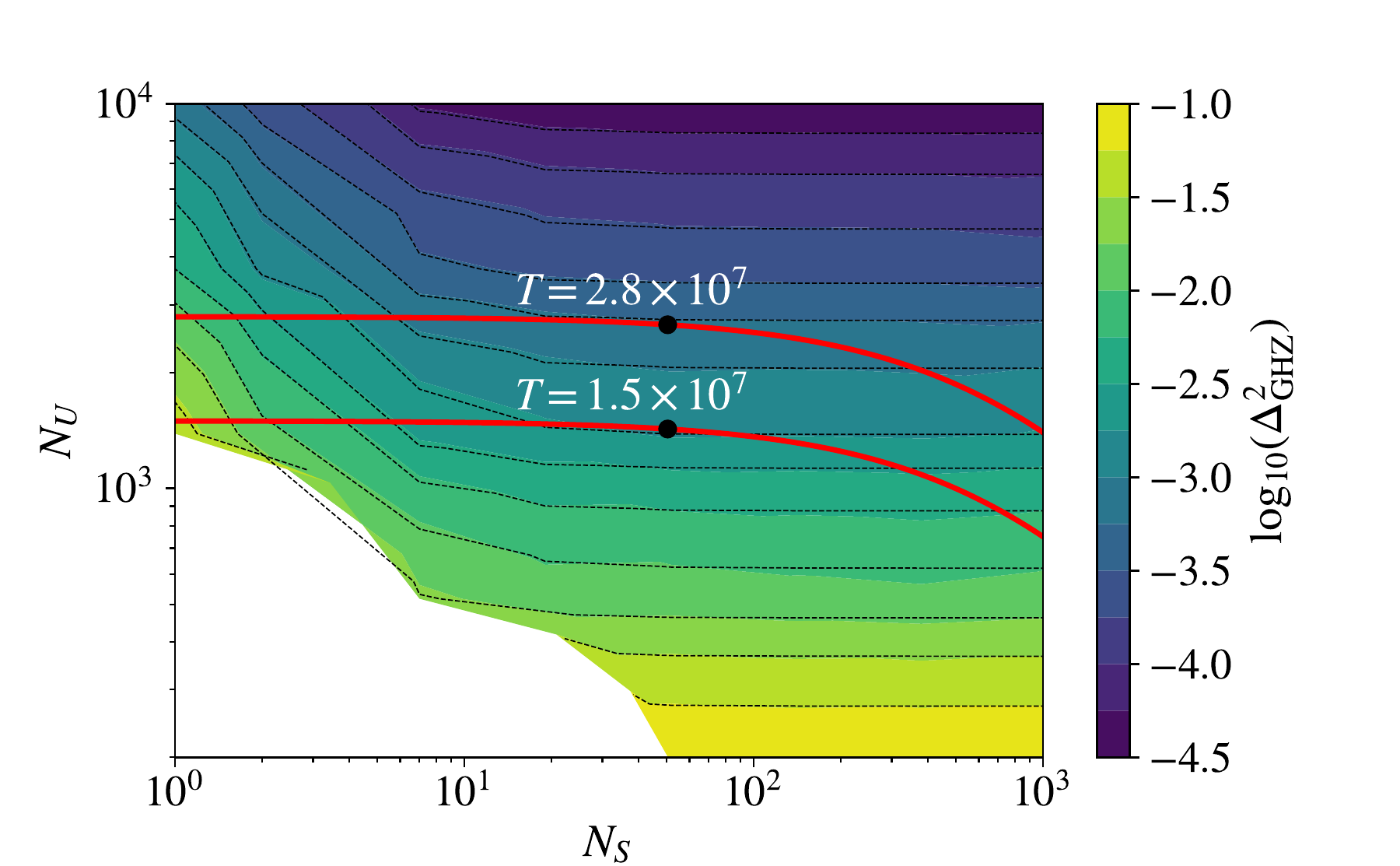}
    \caption{ Contour plot of the simulated error mitigation for the GHZ state for varying $N_S$ and $N_U$. The color bar indicates logarithm of the mean squared error $\Delta_{\rm{GHZ}}^2$.  The dashed lines indicate the contours obtained from the empirical fit $\Delta_{\rm{GHZ}}^2 = \frac{3384}{N_U^2}(1+\frac{22}{N_S^2})$. The red lines show the contours of fixed $T=N_U(1000+N_S)$ for $T=1.5\times 10^7$ and $T=2.8\times 10^7$, with circles indicating the optimal choice of $N_U$ and $N_S$. In the white region of the plot $\Delta_{\rm{GHZ}}^2>0.1$. }
    \label{fig:contour}
\end{figure}

To optimize resources, we first repeat our simulations by fixing the state $\ket{\psi_R}$ in Eq.~\eqref{eq:random mixed state} to be a 5-qubit GHZ state and set $\epsilon=0.1$. This allows us to extract the scaling of errors with resources for this particular state. By examining the simulation data we empirically find that the MSE  scales as $\Delta_{\rm{GHZ}}^2 = \frac{3384}{N_U^2}(1+\frac{22}{N_S^2})$, which is better than the average scaling observed previously (see Supplementary Material). In Fig.~\ref{fig:contour} we compare our empirical fit with the numerically obtained contour and find a good agreement between the two. Next, we model the experiment time by $T=N_U(1000+N_S)$ to capture the trade-off between changing the measurement basis and repeating the measurements in the same basis. Finally, for a fixed $T$ we find the optimal choice of $N_S$ and $N_U$ that gives us the lowest error (see Fig.~\ref{fig:contour}). We note that the optimal choice of $N_S$ and $N_U$ obtained in our simulation may not be the optimal choice for the experiment, as their values may depend on the specific error channel and the purity of the experimental state. Nevertheless, it can serve as a heuristic for better allocating resources in an experiment. 

\begin{figure}
    \includegraphics[width=\columnwidth]{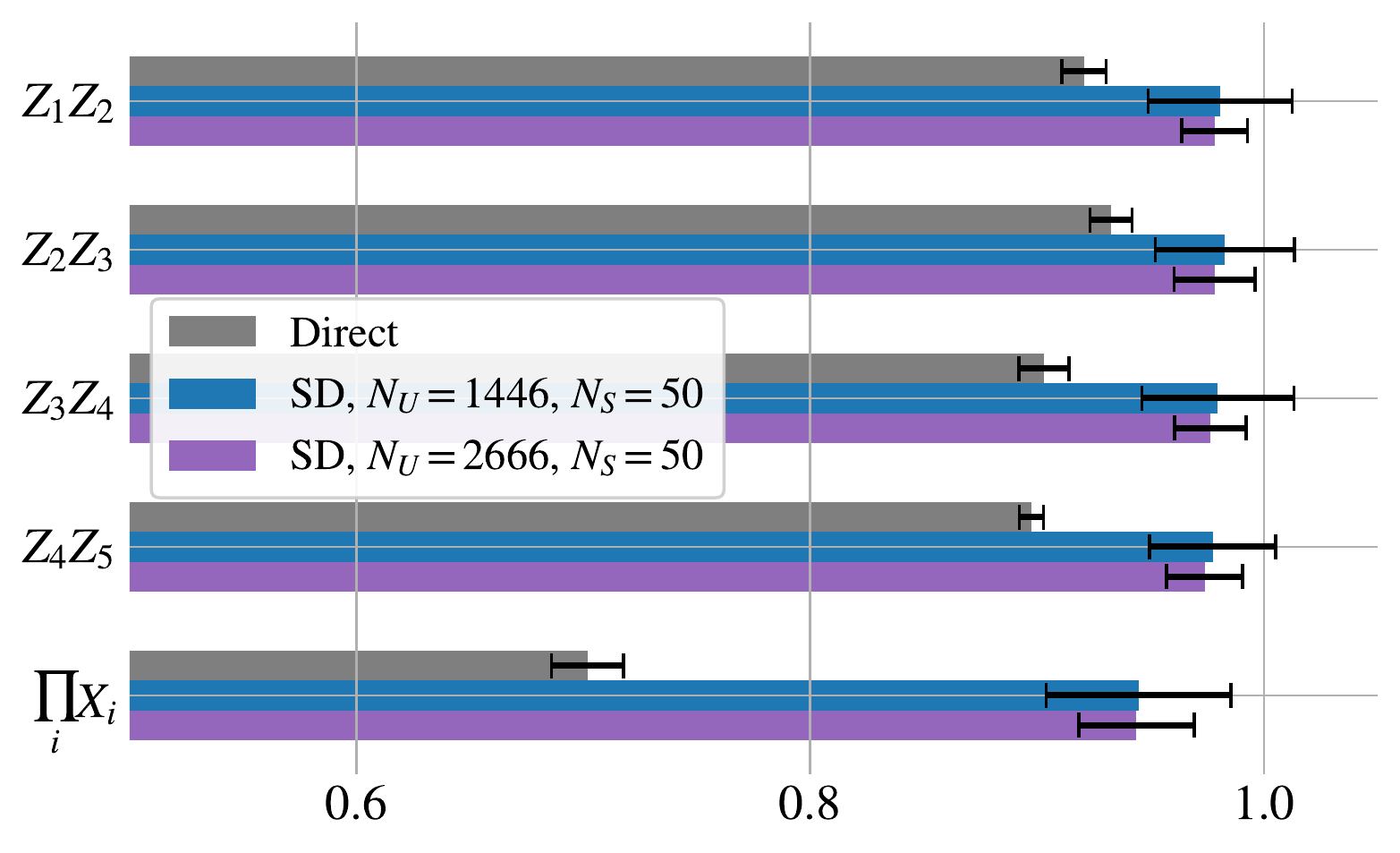}
    \caption{ We compare the experimental results of directly measured expectation values $\tr(\rho O)$ with mitigated values $\expval{O}_{2}$ using $N_U=1446$ and $N_U=2666$ and $N_S=50$ measurements. The labels on the y-axis indicate the choices for $O$ and the x-axis shows the expectation value. Increasing $N_U$ reduces the error on the estimate.  Error bars are standard deviation obtained from bootstrap sampling.}
    \label{fig:experiment}
\end{figure}

After finding the optimal choice of $N_U$ and $N_S$ we resample the experimental measurement data of Ref.~\cite{zhu2021cross} and 
use our error mitigation scheme to recover the expectation values of the stabilizers.  Specifically, in Fig.~\ref{fig:experiment} we observe that $\prod_i X_i $, which is the operator that is most severely affected by the errors benefits the most from the SD scheme. In Supplementary Material we simulate and analyze possible sources of errors in the experiment and based on the performance of SD identify detection errors and dephasing as major sources of noise in the system. Moreover, by increasing $N_U$ from 1446 to 2666 corresponding to the optimal choice for $T=1.5\times10^7$ and $2.8\times10^7$, respectively (shown in Fig.~\ref{fig:contour}), we observe that the error bars (standard deviation obtained by bootstrap resampling) in the mitigated values decrease (see  Fig.~\ref{fig:experiment}). 

\section{Discussion}
We have shown that it is possible to mitigate state-preparation errors using classical shadows and provided numerical evidence of a better sample complexity of this approach compared to full state tomography. We discussed the possibility of incorporating prior knowledge in our estimates and presented a scheme for optimizing measurement resources given experimental constraints. It is interesting to further develop these heuristics to enhance the capabilities of quantum devices in the near term.  

Another aspect of the resource analysis, in addition to the sample complexity, is classical post-processing. As mentioned earlier, the complexity of evaluating the mitigated expectation values using $M=N_U\times N_S$ snapshots scales as $\mathcal{O}(M^2)$. If our numerical  error scaling persists (see Fig.~\ref{fig:all_scalings}(d)) we expect $M\sim 2^{0.82 n_q}$. Note that the second-order mitigation $(m=2)$ has its limitations and even with infinitely many measurements one cannot completely eliminate the errors. One can obtain the full density matrix by taking the average of the measurement snapshots, which allows mitigation with an arbitrary $m$. Therefore, the ultimate mitigation ($m\to \infty$) can be achieved by obtaining the dominant eigenvector of $\rho$~\cite{Koczor_2021}, which takes the time $\mathcal{O}(2^{3 n_q})$. However, taking the latter approach has the same complexity as simulating the full quantum system and is unlikely to be useful beyond a proof-of-concept illustration. Therefore, we believe that the application of our proposed SD method is at the limit where storing and manipulating the full density matrix is out of reach, but storing the shadows and processing them is possible.

Finally, we note that the data collected for SD do not have to come from a single experimental platform. Combining data from different experiments might help with turning coherent errors into incoherent ones that can be mitigated using this scheme. Such a parallel approach helps mitigate errors when multiple experimental systems are available, but performing coherent operations between those systems is not possible.

\section*{acknowledgments}
We thank Andreas Elben, Hsin-Yuan Huang, and Beno\^it Vermersch for helpful discussions. We thank Norbert Linke for helpful comments  and for sharing data from Ref.~\cite{zhu2021cross} for this work. We gratefully acknowledge  Y. Zhu, A. M. Green, C. Huerta Alderete and N. H. Nguyen who took the measurements. We acknowledge support from the ARO (W911NF-18-1-0020, W911NF-18-1-0212), ARO MURI (W911NF-16-1-0349, W911NF-21-1-0325), AFOSR MURI (FA9550-19-1-0399, FA9550-21-1-0209), AFRL (FA8649-21-P-0781), DoE Q-NEXT, NSF (OMA-1936118, EEC-1941583, OMA-2137642), NTT Research, and the Packard Foundation (2020-71479).
S.Z. acknowledges funding provided by the Institute for Quantum Information and Matter, an NSF Physics Frontiers Center (NSF Grant PHY-1733907). A.S. is supported by a Chicago Prize Postdoctoral Fellowship in Theoretical Quantum Science.  Z.P. is supported by AFOSR FA9550-19-1-0399, ARO W911NF2010232, W911NF-15-1-0397 and NSF Physics Frontier Center at the Joint Quantum Institute.  

\emph{Note added.}
Recently, we became aware of a related work~\cite{hu2022logical} that uses similar techniques for error mitigation.

\bibliographystyle{apsrev}
\bibliography{main}

\onecolumngrid
\appendix
\pagebreak
\section{An unbiased estimator for $\tr(O \rho^2)$}
In this section, we show that our estimator in Eq.~\eqref{eq:shadow_average} is unbiased. We first remind the reader that $\{U_j\}_{j=1}^{N_U}$ denotes the $N_U$ sampled unitary operators from random local Clifford gates, and $\{\ket{b^{(i_j)}}_{i_j=1}^{N_S}\}$ denotes the measurement outcomes of $N_S$ measurements fixing $U=U_j$. We can then expand $\hat\rho_j =\frac{1}{N_S}\sum_{i_j=1}^{N_S} \hat{\rho}_{U_j,b^{(i_j)}}$ in Eq.~\eqref{eq:shadow_average} and calculate its expectation value 
\begin{align}
\mathbb{E}\hat{o}_2&=\frac{1}{N_U (N_U-1)}\frac{1}{N_S^2} \mathbb{E}\sum_{\substack{i,i',j,j' \\ i\neq i'}} {\rm{tr}}[V^{(2)}(O\hat{\rho}_{U_j,b^{(i_j)}}) \otimes \hat{\rho}_{U_{j'},b^{(i'_{j'}}})] \\
&= \frac{1}{N_U (N_U-1)N_S^2}[\underbrace{N_U (N_U-1) N_S (N_S-1){\rm{tr}}(O\rho^2)}_{i\neq i',j\neq j'} + \underbrace{N_U (N_U-1) N_S {\rm{tr}}(O\rho^2)]}_{i\neq i',j=j'} \\
&= \frac{N_U (N_U-1)N_S^2}{N_U (N_U-1)N_S^2} {\rm{tr}}(O\rho^2) \\
&= {\rm{tr}}(O\rho^2),
\end{align}
where we used the identity $\tr(O\rho^2) = \mathbb{E}_{U,b, U', b'} [\tr( V^{(2)} (O\hat{\rho}_{U,b}) \otimes \hat{\rho}_{U', b'})]$ in the second line.

\section{Detail of numerical simulations}
In this section, we provide the details of the numerical simulations performed for the scaling of the errors $\Delta^2$ with $N_U$, $N_S$ and the purity $\tr(\rho^2)$. 
We first generate a random mixed state defined in Eq. \eqref{eq:random mixed state} by sampling a random unitary operator from the Haar distribution. 

In order to generate the mixed state with certain purity $\tr(\rho^2)$, we note that the purity is solely determined by the parameter $\epsilon$ and 
\begin{eqnarray}
\tr(\rho_R^2) = (1-\epsilon)^2 + (\frac{\epsilon}{2^{n_q}-1})^2.
\end{eqnarray}
One can therefore vary the parameter $\epsilon$ to tune the purity of the mixed state.

To estimate the squared error $\Delta_R^2$ defined in Eq.~\eqref{eq:shadow_estimation_error} as function of $N_U$, $N_S$ and $\tr(\rho^2)$, we use the bootstrap resampling technique. We perform randomized measurements for 10000 different random bases, each with 10000 shots. These data form the empirical distribution of the classical shadow for a given $\rho_R$. 

For a given pair of $(N_U, N_S)$, we sample the classical shadow for $N_U$ random basis and $N_S$ shots from the the empirical distribution and estimate $\langle O \rangle_{(2)}$ using Eq.~\eqref{eq:shadow_average}. The squared error of the estimation, $\Delta_R$ is defined as the squared difference between the estimation and the exact value $\langle O \rangle_{(2)}$ as defined in Eq. \eqref{eq:shadow_estimation_error}. We perform the resampling 250 times to obtain the average of $\Delta_R^2$ denoted by $\overline{\Delta_R^2}$. 

Finally, we average over the random mixed states $\rho_R$ by generating $N_R = 100$ different random mixed states $\rho_R$ and calculate $\Delta^2 = \frac{1}{N_R} \sum_R \overline{\Delta_R^2}$. The standard deviation used for plotting the error bars is given by ${\rm{std}}(\Delta^2) = \sqrt{\sum_R \frac{1}{N_R}(\Delta^2 - \overline{\Delta_R^2})^2}$.

\section{Numerical simulations of the GHZ state}
In this section, we provide more information about the simulations of the GHZ state used for producing Fig.~\ref{fig:contour}. 

\begin{figure}
    \centering
    \includegraphics[width=0.6\columnwidth]{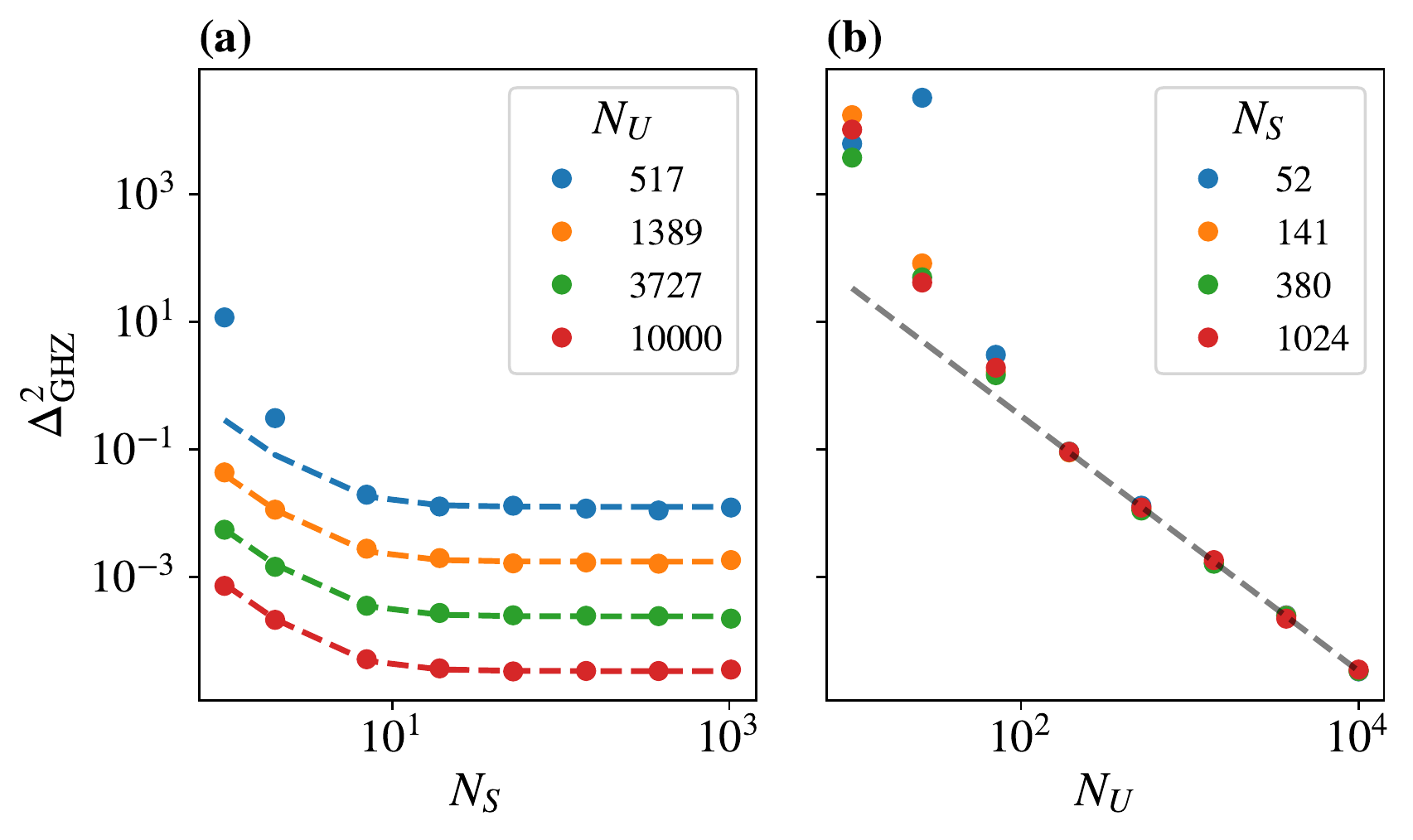}
    \caption{Scaling of the mean squared error $\Delta_{\rm{GHZ}}^2$ for a 5- qubit GHZ state for (a) varying $N_S$ and (b) $N_U$. The dashed lines indicate the empirical fit $\Delta_{\rm{GHZ}}^2 = \frac{3384}{N_U^2}(1+\frac{22}{N_S^2})$.}
    \label{fig:ghz_fit}
\end{figure}

We generate a 5-qubit GHZ state with $\epsilon=0.1$ and for each value of $N_S$ and $N_U$ simulate the randomized measurement protocol 1000 times. We then calculate the mean squared error $\Delta^2_{\rm{GHZ}}$ using these samples, see Fig.~\ref{fig:ghz_fit}. Based on the observed scaling for large $N_S$ and $N_U$, we use the expression $\Delta_{\rm{GHZ}}^2 = \frac{c_1}{N_U^2}(1+\frac{c_2}{N_S^2})$ to fit the data and find that $c_1=3384$ and $c2=22$. Since the values $\Delta^2_{\rm{GHZ}}$ span orders of magnitude, we use $\log_{10}(\Delta^2_{\rm{GHZ}})$ to fit the data capture the correct behavior across a large range of values. 

\section{A biased estimator for purity}
As noted in the main text, it might be beneficial to incorporate prior knowledge about the value of the purity $\tr(\rho^2)$ to reduce the errors. We now show one approach to incorporating prior knowledge using a Gaussian prior and Bayes' rule. 

Let $\mu$ denote the true value of $\tr(\rho^2)$ and assume that we have a prior belief that $\mu\sim N(\mu_0,\sigma_0^2)$, i.e., a normal distribution with the mean $\mu_0$ and the variance $\sigma_0^2$. Next, assume that after performing an experiment we estimate the purity to be $s_2$. We also assume that this observation is normally distributed with the variance, $\sigma^2$, that is known. Therefore, based on our measurements and assumptions we have 
\begin{equation}
    \prob(s_2|\mu)=\frac{1}{\sqrt{2\pi \sigma^2}}\exp[-\frac{(s_2-\mu)^2}{2\sigma^2}].
\end{equation}
Moreover, our prior is 
\begin{equation}
    \prob(\mu)=\frac{1}{\sqrt{2\pi \sigma_0^2}}\exp[-\frac{(\mu-\mu_0)^2}{2\sigma_0^2}].
\end{equation}
Then using Bayes' rule $    \prob(\mu|s_2)=\frac{\prob(\mu)\prob(s_2|\mu)}{\prob(s_2)}$ we find the posterior 
\begin{align}
    \prob(\mu|s_2)&\propto \prob(\mu)\prob(s_2|\mu)\\
    &=\exp[-\frac{(s_2-\mu)^2}{2\sigma^2}-\frac{(\mu-\mu_0)^2}{2\sigma_0^2}]\\
    &\propto\exp[-\frac{(\mu-\mu')^2}{2\sigma'^2}],
\end{align}
where our updated mean and variance are
\begin{align}
\mu' &= \frac{\mu_0 \sigma^2+ s_2  \sigma_0^2}{\sigma^2+\sigma_0^2},\\
\sigma'^2 &= \frac{\sigma^2 \sigma_0^2}{\sigma^2+\sigma_0^2}. 
\end{align}
We now use $\mu'$ as our estimator for purity. We assume that $\sigma^2\propto 1/N_U$, and define a parameter $\alpha$ such that $\frac{\sigma^2}{\sigma_0^2}=\frac{\alpha}{N_U}$. We then have
\begin{equation}\label{eq:purity_biased}
    \mu' = \frac{s_2 + \alpha \frac{\mu_0}{N_U}}{1+\frac{\alpha}{N_U}}.
\end{equation}
We can then treat $\alpha$ as a hyperparameter that quantifies our confidence in our initial guess. Large values of $\alpha$ indicate our high confidence in $\mu_0$. 

\begin{figure}
    \centering
    \includegraphics[width=0.6\columnwidth]{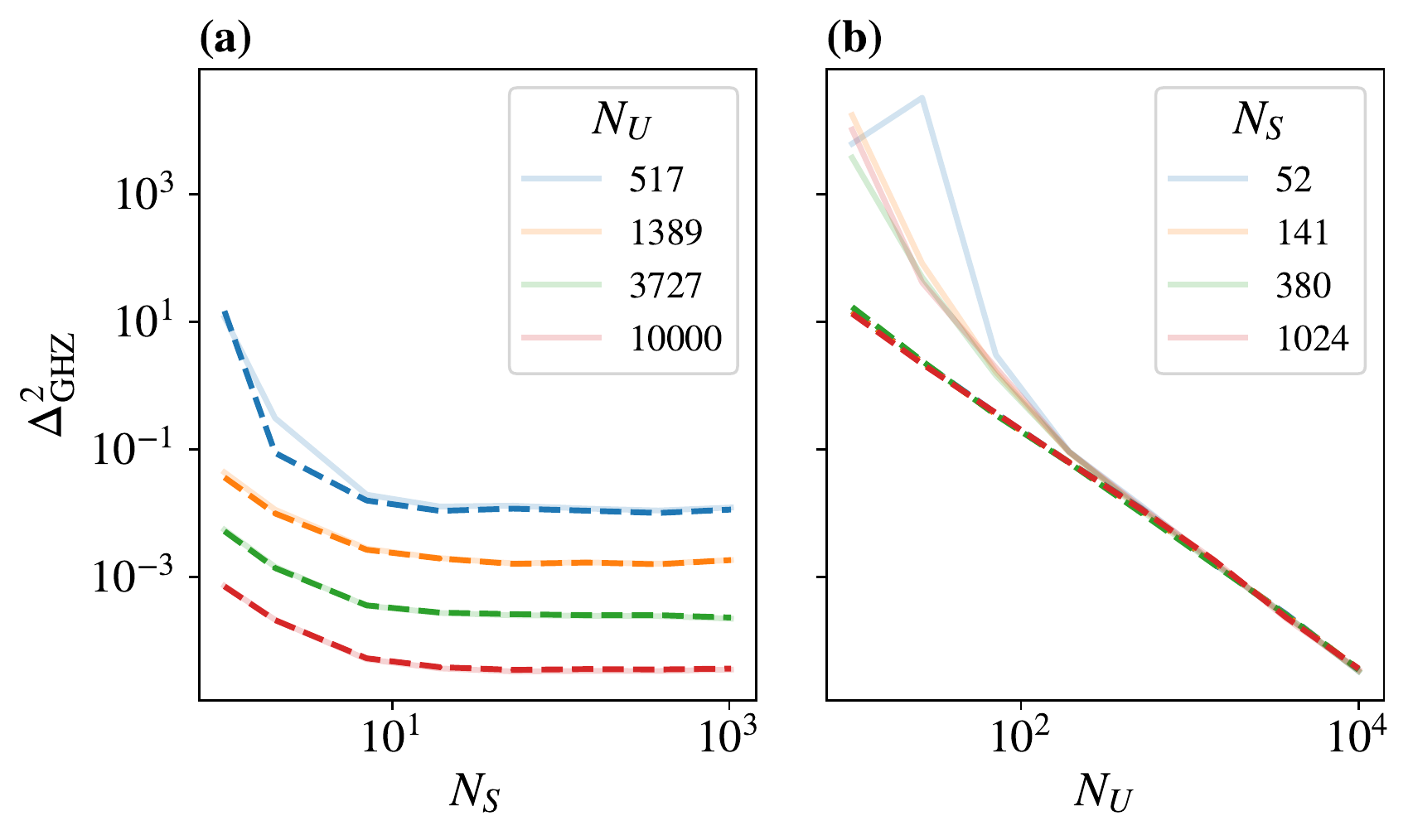}
    \caption{Using the biased estimator for purity in Eq.~\eqref{eq:purity_biased} on the data from Fig.~\ref{fig:ghz_fit} reduces the errors for smaller values of $N_U$.  The dashed lines show the errors calculated using the biased estimator, where the solid lines are the original data from Fig.~\ref{fig:ghz_fit}.}
    \label{fig:ghz_biased}
\end{figure}

This method is particularly useful if we have a good guess about the purity of the state in our experiment. To illustrate, we apply this modified estimator to our data in Fig.~\ref{fig:ghz_fit}, with $\mu_0=0.9$ and $\alpha=100$. The true value of purity in this case 0.81. The results in Fig.~\ref{fig:ghz_biased} show that, even with more than 10\% error in the prior, using this biased estimator improves the errors for smaller values of $N_U$. 

\section{Details of the experiment and errors}
\subsection{Experimental setup}
    
\begin{figure}
        \centering
        \includegraphics[width=\columnwidth]{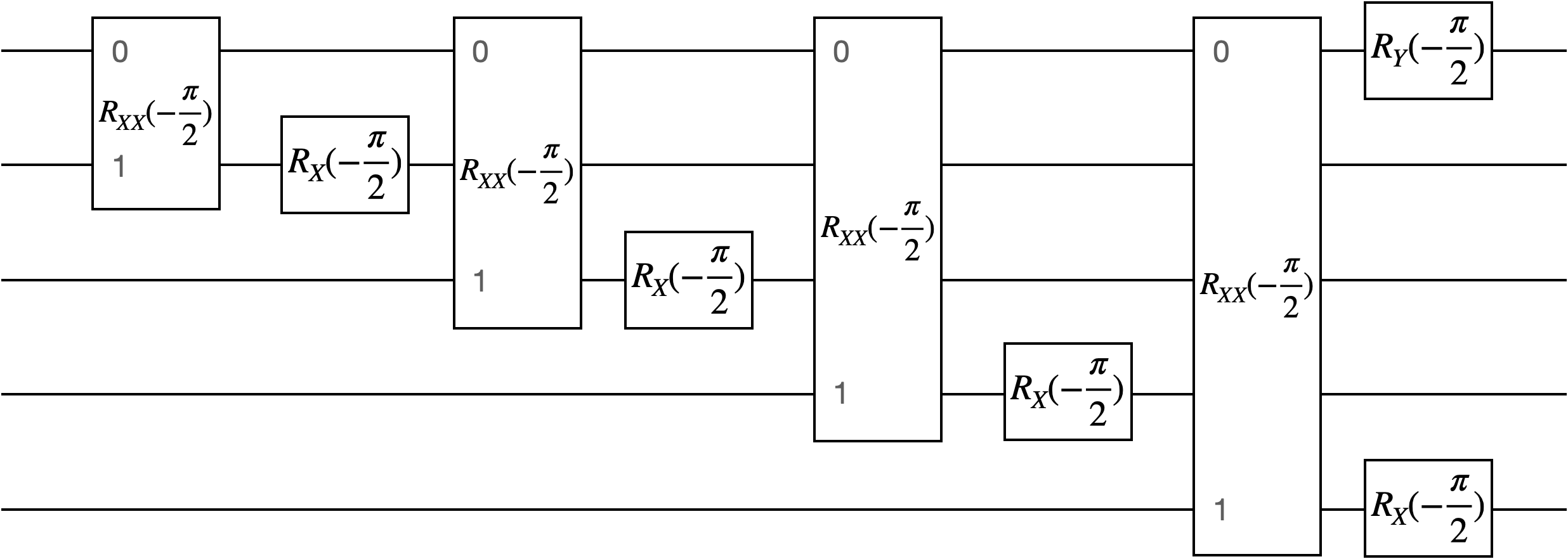}
        \caption{The circuit used for preparing the GHZ state in the trapped ion quantum computer, where $R_{XX}(\theta) = \exp(-i \frac{\theta}{2} XX)$, $R_\alpha(\theta) = \exp(-i \sigma_\alpha \frac{\theta}{2})$ and $\alpha = x, y, z$.}
        \label{fig:ghz}
    \end{figure}

The trapped-ion experiment is performed on a quantum computer consisting of a chain of nine 171Yb$^+$ ions confined in a Paul trap with blade electrodes.  Typical single- and two-qubit gate fidelities are $99.5(2)\%$ and $98-99\%$.  Detailed performance of the system is described in Ref. \cite{landsman2019verified}.
 The GHZ state in the experiment is prepared by running the circuit show in Fig.~\ref{fig:ghz} on five qubits. The circuit utilizes the two-qubit gate $R_{XX}(\theta) = \exp(-i \frac{\theta}{2} XX)$, and the single qubit rotations $R_\alpha(\theta) = \exp(-i \sigma_\alpha \frac{\theta}{2})$ with $\alpha = x, y, z$.
    
\subsection{ Error channels and simulation}
    
In this section, we describe the detailed implementations for the simulation of the error channels. We simulate the circuit in Fig.~\ref{fig:ghz} on a classical computer. 

To simulate coherent errors, we replace the $R_{XX}(\theta)$ gate by $R_{XX}(\theta(1 + \delta_{\rm{coh}}))$, where $\delta_{\rm{coh}}$ is the over-rotation rate. 

The dephasing error is simulated by applying the following noise channel at the end of the simulation:
\begin{eqnarray}
   \mathcal{C}(\rho) = (1-p_{\rm{deph}}) \rho + \frac{p_{\rm{deph}}}{n_q} \sum_{i = 1}^{n_q} Z_i  \rho Z_i,
\end{eqnarray}
where $p_{\rm{deph}}$ is the dephasing rate.

To simulate detection errors in the measurements we first rotate the density matrix to the basis that the measurement will be performed. For example, to measure the $\prod_{i=1}^{n_q} X_i$ operator, we perform Hadamard rotation for all the qubits. After the rotation, we take the diagonal part of the density matrix, $P$. It corresponds to the probability distribution of the measurement outcomes. We then apply the detection error matrix, $M$, to the probability distribution $P$. In this work, we focus on  uncorrelated detection errors. The matrix $M$ in this case is given by 
\begin{eqnarray}\label{eq:spammatrix}
   M = \bigotimes_{i = 1}^{n_q} A_i,
\end{eqnarray}
where 
\begin{eqnarray}
   A_i = \begin{pmatrix}
   1-p_{0} & p_1\\
   p_0 & 1-p_1
   \end{pmatrix},
\end{eqnarray}
where $p_0 (p_1)$ denotes the  probability that the detector gives outcome $1(0)$ where the true outcome should be $0(1)$, respectively. We assume that $p_0 = p_1 = p_{\rm{det}}$ for simplicity. After the application of $M$, we calculate the expectation value of the observables based on the modified probability distribution. 

To simulate the measurement of second order mitigation $\langle O \rangle _{(2)}$ with detection errors, we first simulate the measurement of all $4^n$ Pauli strings with detection errors using the method described in the previous section. We then define the reconstructed density matrix as $\rho = \frac{1}{2^{n_q}} \sum_{k = 0}^{4^{nq}-1} c_k P_k$, where $P_k$ is the $k$th Pauli string operator and $c_k$ is the simulated measurement result of $P_k$ with detection errors. Finally, the second order mitigation is computed as $\langle O \rangle_{(2)} = \frac{\tr(O \rho^2)}{\tr(\rho^2)}$.

\subsection{Analysis of errors}

\begin{figure}
     \includegraphics[width=0.6\columnwidth]{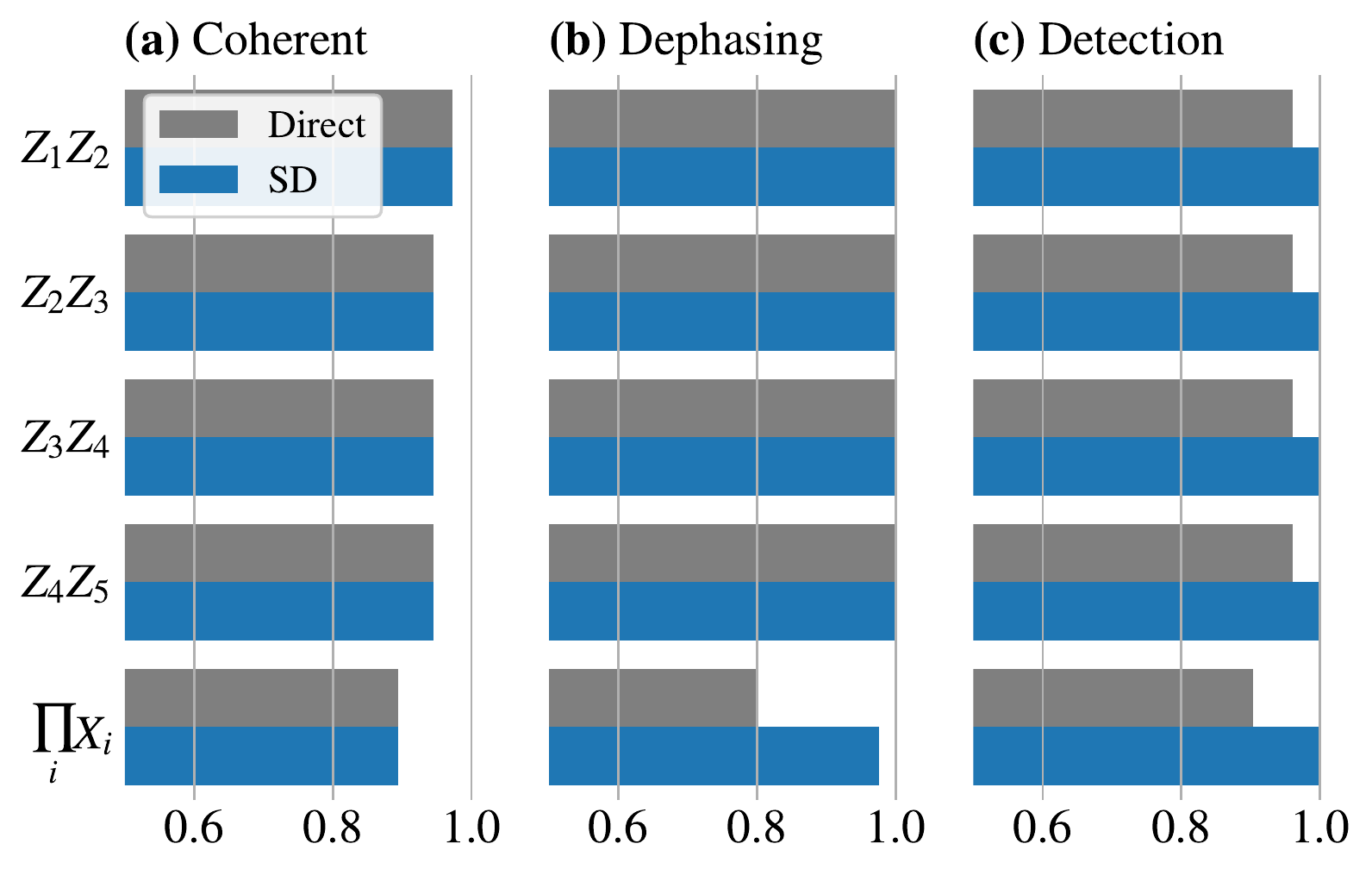}
    \caption{Simulated effect of errors on second order mitigation. We compare the direct approach $\tr(\rho O)$ (Direct) against the second order mitigation $\expval{O}_{(2)}$ (SD) for (a) coherent error with $\delta_{\rm{coh}} = 0.15$ relative over-rotation, (b) single qubit dephasing error with $p_{\rm{deph}} = 0.1$, and detection errors with $p_{\rm{det}}=0.01$}. 
    \label{fig:simulation}
\end{figure}

In addition to correcting the expectation values, our method also reveals some facts about the nature of errors in the system. We first note that static coherent errors do not benefit from SD (see Fig.~\ref{fig:simulation}(a)). This is because these errors change the eigenstates of $\rho$ while leaving the eigenvalues unaffected.  From the experimental results in Fig.~\ref{fig:experiment}, we can see that $\prod_i {X_i}$ is the operator that is most affected by the errors. However, the fact that it benefits considerably from the error mitigation protocol suggests that the errors are mostly incoherent. These observations are further validated by the numerical simulation of coherent errors (Fig.~\ref{fig:simulation}(a)), dephasing errors (Fig.~\ref{fig:simulation}(b)), and detection  errors (Fig.~\ref{fig:simulation}(c)). We see that, unlike coherent errors, the latter two benefit from SD. The contrast between the $ZZ$ and $\prod_i X_i$ can be due to either dephasing or detection errors. However, in the next section, we provide a  detailed analysis using a different error mitigation technique~\cite{shen2012correcting} that only mitigates detection errors, and show that it is unlikely that detection errors are the only source of errors in this experiment.  The residual errors in Fig.~\ref{fig:experiment} either correspond to higher-order incoherent errors, incoherent errors that modify the eigenvectors of $\rho$ (also known as the coherent mismatch~\cite{koczor2021exponential,huggins2020virtual,Koczor_2021}), or coherent errors originating from under(over)-rotation in two-qubit gates, which is a known source of error in trapped-ion systems~\cite{maksymov2021detecting}.

\subsection{Correcting detection errors}
\begin{figure}[h]
    \centering
    \includegraphics[width=0.6\columnwidth]{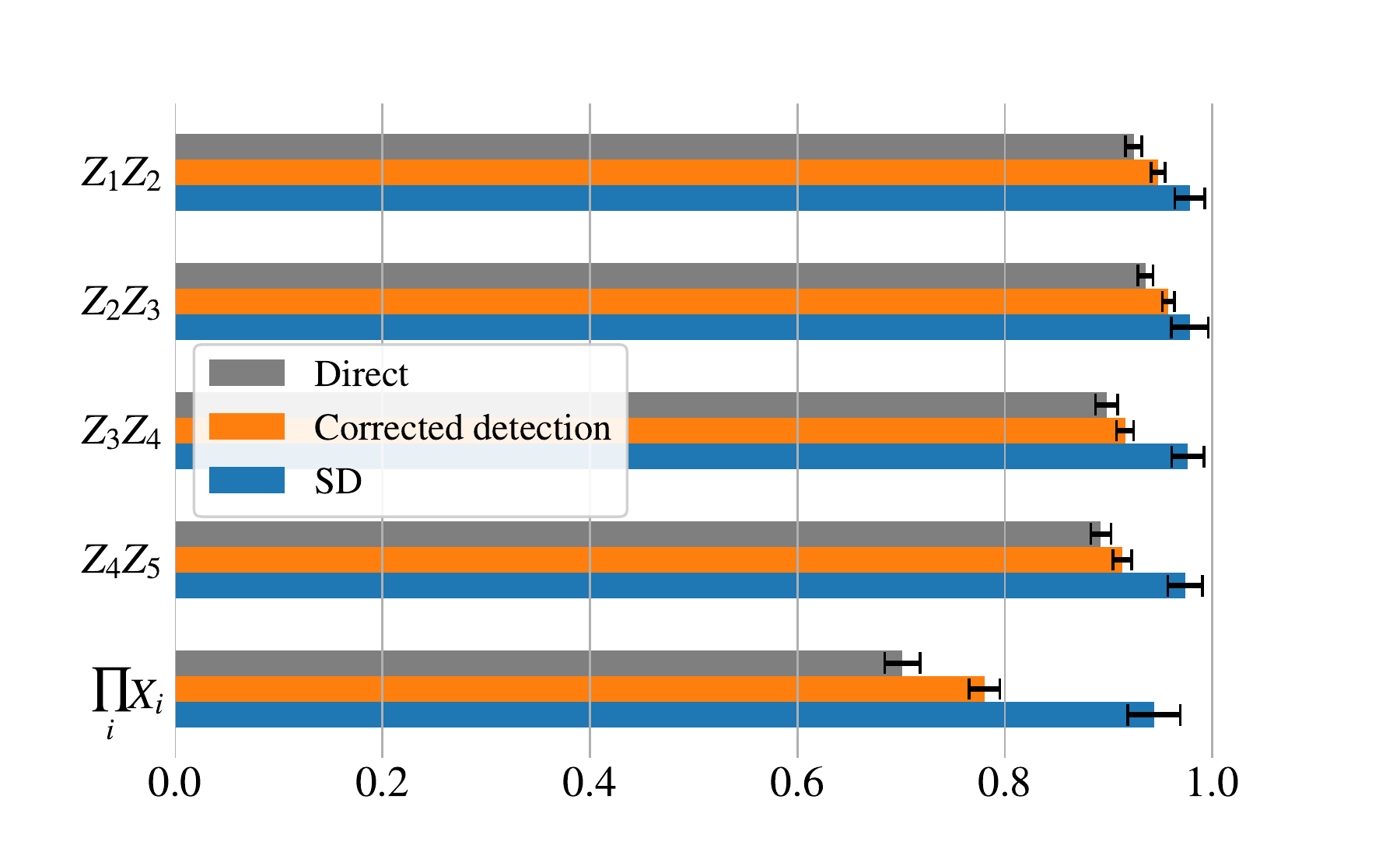}
    \caption{Comparison of the experimental data (Direct), corrected data using detection calibration (Calibrated detection), and shadows distillation (SD) with $N_U=2666$ and $N_S=50$  measurements. The gap between the corrected detection and SD data in $\expval{\prod_i X_i}$ and its absence in $\expval{ZZ}$s can be explained by the existence of dephasing error in the experiment. }
    \label{fig:spam}
\end{figure}

It is also possible to correct detection errors by first calibrating the matrix $M$~\eqref{eq:spammatrix} in the experiment and applying $M^{-1}$ to the vector of outcome probabilities obtained from the measurements~\cite{shen2012correcting}. We apply this procedure to the experimentally obtained expectation values and show the results in Fig.~\ref{fig:spam}. We observe that the corrected expectation values are still lower than those obtained from SD, which indicates that SD is mitigating errors beyond just those in the detection process. 

\section{Analytical upper bounds on the estimation variance as a function of \texorpdfstring{$N_S$}{N\_S} and \texorpdfstring{$N_U$}{N\_U}}

\subsection{Definition}

Given an $n$-qubit ($n=n_q$ in the main text) quantum state $\rho$, we perform a random local Clifford unitary $U$ operation on $\rho$ and then perform the computational basis measurement $N_S$ times. Suppose $\{\ket{b^{(i)}}\}_{i=1}^{N_S}$ are the measurement outcomes (note that here $b^{(i)}$ are $n$-bit strings), then in this section, we define the following unbiased estimator of $\rho$: 
\[
\hat\rho =  \frac{1}{N_S}\sum_{i=1}^{N_S} \hat{\rho}_{U,b^{(i)}} =  \frac{1}{N_S}\sum_{i=1}^{N_S} \mM^{-1}(U^\dagger \ket{b^{(i)}}\bra{b^{(i)}} U), 
\]
where we define
\[
\begin{split}
\mM(\rho) &= \frac{1}{N_S}\sum_{i=1}^{N_S} \bE\left[  U^\dagger \ket{b^{(i)}}\bra{b^{(i)}} U \right]= \frac{1}{N_S}\sum_{i=1}^{N_S} \bE_{U \sim \mU}\sum_{\{b^{(i)}\}} \left[  U^\dagger \ket{b^{(i)}}\braket{b^{(i)}|U \rho U^\dagger|b^{(i)}}\bra{b^{(i)}} U \right] = \mD_{1/3}^{\otimes n}(\rho),
\end{split}
\]
\[
\mM^{-1}(\rho) = (\mD_{1/3}^{-1})^{\otimes n}(\rho),
\]
where $\mD_{1/3}(\rho) = \frac{1}{3}\rho + \frac{1}{3}\trace(\rho) {{I}}$, $\mD_{1/3}^{-1}(\rho) = 3\rho - \trace(\rho) {{I}}$ and $\mU$ denotes the uniform distribution of local Clifford operations on $n$ qubits. 

\subsection{Variance of estimating \texorpdfstring{$\trace(O\rho)$}{tr(O*rho)} }

Clearly, $\trace(O\hat\rho)$ is an unbiased estimator of $\trace(O\rho)$. Now we compute its variance. 
\[
\begin{split}
&\quad\; \bV[\trace(O\hat\rho)] \\
&= \bE_{U\sim\mU} \sum_{\{b^{(i)}\}} \left(\prod_{i=1}^{N_S} \braket{b^{(i)}|U\rho U^\dagger|b^{(i)}} \right)\left(\frac{1}{N_S}\sum_{i=1}^{N_S} \braket{b^{(i)}|U \mM^{-1}(O) U^\dagger|b^{(i)}}\right)^2 - \trace(\rho O)^2 \\
&= \frac{1}{N_S^2}\bE_{U\sim\mU} \sum_{\{b^{(i)}\}} \left(\prod_{i=1}^{N_S} \braket{b^{(i)}|U\rho U^\dagger|b^{(i)}} \right)\sum_{i,i'=1}^{N_S} \braket{b^{(i)}|U \mM^{-1}(O) U^\dagger|b^{(i)}}\braket{b^{(i')}|U \mM^{-1}(O) U^\dagger|b^{(i')}} - \trace(\rho O)^2\\
&= \frac{1}{N_S}\left( \bE_{U\sim\mU} \sum_{b}  \braket{b|U\rho U^\dagger|b} \braket{b|U \mM^{-1}(O) U^\dagger|b}^2  - \trace(\rho O)^2\right) +  \\
&\quad \frac{N_S-1}{N_S}\left( \bE_{U\sim\mU} \sum_{b,b'}  \braket{b|U\rho U^\dagger|b}\braket{b'|U\rho U^\dagger|b'} \braket{b|U \mM^{-1}(O) U^\dagger|b}\braket{b'|U \mM^{-1}(O) U^\dagger|b'} - \trace(\rho O)^2\right).
\end{split}
\]

It is known from Proposition S3 in \cite{huang2020predicting} that when $O$ is a weight-$k$ operator and has a Pauli decomposition $O = \sum_{\vp} \alpha_\vp P_\vp$, $\vp \in \{I,X,Y,Z\}^{n}$, the first term is equal to 
\[
\begin{split}
&\quad \frac{1}{N_S}\bE_{U\sim\mU} \sum_{b}  \braket{b|U\rho U^\dagger|b} \braket{b|U \mM^{-1}(O) U^\dagger|b}^2  - \frac{1}{N_S}\trace(\rho O)^2 = \frac{1}{N_S}\left( \frac{1}{3^k} \sum_{\vs\in\{X,Y,Z\}^k} \trace(\rho O_\vs^2) - \trace(\rho O)^2\right),
\end{split}
\]
where 
$O_\vs = \sum_{\vq \triangleright \vs} 3^{\abs{\vq}}\alpha_{\vq} P_{\vq}$,
and $\vq \triangleright \vs$ means $q_i$ is equal to either $s_i$ or $I$ for all $i$.

Now we compute the second term.
We first compute  
\[
\begin{split}
\bE_{U\sim\mU} \sum_{b,b'}  (U\otimes U)\ket{bb'}\bra{bb'}(U^\dagger \otimes U^\dagger) \braket{b|U^\dagger \mM^{-1}(P_\vp) U|b}\braket{b'|U^\dagger \mM^{-1}(P_\vq) U|b'} = \bigotimes_{i=1}^n F(p_i,q_i),
\end{split}
\]
where $P_\vq$, $P_\vq$ are Pauli operators.
\[
F(p_i,q_i) = 
\bE_{U_1\in\mU_1} \sum_{x_1,x_2=0}^1 (U_1\otimes U_1)\ket{x_1x_2}\bra{x_1x_2}(U_1^\dagger \otimes U_1^\dagger) \braket{x_1|U_1^\dagger P_{p_i} U_1|x_1}\braket{x_2|U_1^\dagger P_{q_i} U_1|x_2},
\]
where $\mU_1$ is the uniform distribution of Clifford gates on one qubit. 
After a few calculations, we get 
\[
F(p_i,q_i) = \begin{cases}
{{I}} \otimes {{I}} & p_i=q_i=0,\\
\frac{1}{3} P_{p_i} \otimes P_{q_i} & p_i=q_i \neq 0,~ p_i=0,q_i\neq0, \text{~or~} p_i\neq0,q_i=0,\\
0 & \text{otherwise.}
\end{cases}
\]

Let 
\[
f(\vp,\vq) = 
\begin{cases}
0 & \exists i,~\text{s.t.}~p_i\neq q_i\text{~and~}p_i,q_i\neq {{I}}\\
3^s & s = \abs{\{i:p_i=q_i,p_i\neq {{I}}\}},
\end{cases}
\] 
then the second term is equal to $\frac{N_S-1}{N_S} \times $
\[
\begin{split}
&\quad \; \bE_{U\sim\mU} \sum_{b,b'}  \braket{b|U\rho U^\dagger|b}\braket{b'|U\rho U^\dagger|b'} \braket{b|U \mM^{-1}(O) U^\dagger|b}\braket{b'|U \mM^{-1}(O) U^\dagger|b'} - \trace(\rho O)^2\\
&= \trace\left((\rho\otimes \rho)\sum_{\vp\vq} \alpha_\vp \alpha_\vq \bigotimes_{i=1}^n F(p_i,q_i)\right) - \trace(\rho O)^2 =\sum_{\vp\vq} \alpha_\vp \alpha_\vq f(\vp,\vq) \trace(\rho P_\vp)\trace(\rho P_\vq) - \trace(\rho O)^2\\
&= \left( \frac{1}{3^k} \sum_{\vs\in\{X,Y,Z\}^k} \trace(\rho O_\vs)^2 - \trace(\rho O)^2\right)
\end{split}
\]

Therefore, we have 
\begin{align}
\bV[\trace(O\hat\rho)] &=  \frac{1}{N_S}\left( \frac{1}{3^k} \sum_{\vs\in\{X,Y,Z\}^k} \trace(\rho O_\vs^2) - \trace(\rho O)^2\right) + \frac{N_S-1}{N_S} \left( \frac{1}{3^k} \sum_{\vs\in\{X,Y,Z\}^k} \trace(\rho O_\vs)^2 - \trace(\rho O)^2\right)\nonumber\\
&=: u_0(O,\rho) + \frac{1}{N_S} u_1(O,\rho),
\end{align}
where 
\begin{equation}u_0(O,\rho) := \frac{1}{3^k} \sum_{\vs\in\{X,Y,Z\}^k} \trace(\rho O_\vs)^2 - \trace(\rho O)^2\text{~~and~~} 
u_1(O,\rho) := \frac{1}{3^k} \sum_{\vs\in\{X,Y,Z\}^k} \trace(\rho O_\vs^2) - \trace(\rho O_\vs)^2
\end{equation}
are all non-negative functions of $O$ and $\rho$. It is known from \cite{huang2020predicting} that 
\[
u_0(O,\rho) + u_1(O,\rho) \leq 2^k \trace(O^2). 
\]
Clearly, a large $N_S$ is helpful as long as $u_0(O,\rho)$ is significantly smaller than $u_1(O,\rho)$.

\subsection{Variance of estimating \texorpdfstring{$\trace(O\rho^2)$}{tr(O*rho\^{}2)}}

Let $\hat o_2 = \frac{1}{N_U(N_U-1)}\sum_{j\neq j'}\trace( {V^{(2)}} \hat\rho_j \otimes (O\hat\rho_{j'}))$. This is an unbiased estimator of $o_2 = \trace(O\rho^2)$. Now we compute its variance.
Let ${O^{(2)}} := \frac{1}{2}({V^{(2)}}({{I}}\otimes O)+({{I}} \otimes O){V^{(2)}})$, we have 
\[
\hat{o}_2 = \binom{N_U}{2}^{-1} \sum_{j < j'} \trace({O^{(2)}} \hat{\rho}_j \otimes \hat{\rho}_{j'}),
\]
where $\hat\rho_j = \frac{1}{N_S}\sum_{i_j=1}^{N_S} \mM^{-1}(U_j^\dagger \ket{b^{(i_j)}}\bra{b^{(i_j)}} U_j)$, $\{U_j\}_{j=1}^{N_U}$ is sampled from random local Clifford gates, and $\{\ket{b^{(i_j)}}_{i_j=1}^{N_S}\}$ are measurement outcomes of $N_S$ measurements fixing $U=U_j$. A total $N_S\cdot N_U$ number of measurements are performed. 

In order to derive an upper bound of $\bV[\hat{o}_2]$, we first note from our discussion above that 
\[
\bV[\trace(A \hat \rho)] \leq u_0(A,\rho) + \frac{1}{N_S} u_1(A,\rho),
\]
for an arbitrary Hermitian operator $A$, and for $j\neq j'$,
\[
\bV[\trace({O^{(2)}} \hat \rho_j \otimes \hat \rho_{j'})] \leq u_0(O^{(2)},\rho\otimes\rho) +  \frac{1}{N_S} u_1(O^{(2)},\rho\otimes\rho). 
\]
Consider
\[
(\hat o_2)^2 = \binom{N_U}{2}^{-2} \sum_{j < j'}\sum_{k < k'} \trace({O^{(2)}} \hat{\rho}_j \otimes \hat{\rho}_{j'})\trace({O^{(2)}} \hat{\rho}_k \otimes \hat{\rho}_{k'}).
\]
For terms where all indices are distinct, 
\[
\bE[\trace({O^{(2)}} \hat{\rho}_j \otimes \hat{\rho}_{j'})\trace({O^{(2)}} \hat{\rho}_k \otimes \hat{\rho}_{k'})] = \trace(\rho^2 O)^2;
\]
for terms where two of the indices coincide, 
\[
\bE[\trace({O^{(2)}} \hat{\rho}_j \otimes \hat{\rho}_{j'})\trace({O^{(2)}} \hat{\rho}_k \otimes \hat{\rho}_{k'})] = \bE[\trace((\hat\rho \otimes \rho){O^{(2)}})^2] = \bE[\trace(\hat\rho A)^2],
\]
where $A := \frac{1}{2}(\rho O + O \rho)$; for terms where $(j,j')$ coincides with $(k,k')$,
\[
\bE[\trace({O^{(2)}} \hat{\rho}_j \otimes \hat{\rho}_{j'})\trace({O^{(2)}} \hat{\rho}_k \otimes \hat{\rho}_{k'})] = 
\bE[\trace({O^{(2)}} \hat{\rho}_j \otimes \hat{\rho}_{j'})^2]. 
\]
Then we have
\begin{align}
\bV[\hat o_2]&= \binom{N_U}{2}^{-1}\left(2(N_U-1) \left(u_0(A,\rho)+\frac{1}{N_S} u_1(A,\rho)\right) +  \left(u_0(O^{(2)},\rho\otimes\rho) + \frac{1}{N_S} u_1(O^{(2)},\rho\otimes\rho)\right) \right)\nonumber\\
&\leq \frac{4}{N_U} \left(u_0(A,\rho)+\frac{1}{N_S} u_1(A,\rho)\right) + \frac{4}{N_U^2} \left(u_0(O^{(2)},\rho\otimes\rho) + \frac{1}{N_S} u_1(O^{(2)},\rho\otimes\rho)\right).
\end{align}
In particular, when $O = I$, we have 
\begin{align}
\bV[\hat s_2]&= \binom{N_U}{2}^{-1}\left(2(N_U-1) \left(u_0(\rho,\rho)+\frac{1}{N_S} u_1(\rho,\rho)\right) +  \left(u_0(V^{(2)},\rho\otimes\rho) + \frac{1}{N_S} u_1(V^{(2)},\rho\otimes\rho)\right) \right)\nonumber\\
&\leq \frac{4}{N_U} \left(u_0(\rho,\rho)+\frac{1}{N_S} u_1(\rho,\rho)\right) + \frac{4}{N_U^2} \left(u_0(V^{(2)},\rho\otimes\rho) + \frac{1}{N_S} u_1(V^{(2)},\rho\otimes\rho)\right).
\end{align}
\end{document}